\documentclass[12pt]{iopart}

\begin{document}

\title[Benini, Kirillov and Montani]{Oscillatory regime in the Multidimensional Homogeneous Cosmological Models
Induced by a Vector Field}

\author{R. Benini\dag\ , A.A. Kirillov\ddag\  and Giovanni Montani\opendiamond\ }
%\email{riccardobenini@virgilio.it}
\address{\dag\ Dipartimento di Fisica - Universit\`a di Bologna and INFN \\ Sezione di Bologna,
via Irnerio 46, 40126 Bologna, Italy}

%\author{A.A. Kirillov}
%\email{kirillov@unn.ac.ru}
\address{\ddag\ Institute for Applied Mathematics and Cybernetics\\ 
10 Ulyanova str., Nizhny Novgorod, 603005, Russia\\}

%\author{Giovanni Montani}
\address{\opendiamond\ Dipartimento di Fisica (G9)
  and  International Center for Relativistic Astrophysics, %\\
  Universit\`a di Roma I--00185 Roma, Italy. 
}

\date{\today}

%%%%%%%%%%%%%%%%%%%%%%%%%%%%%%%%%%%%%%%%%%%%%%%%%%%%

\begin{abstract}
We show that in multidimensional gravity vector fields
completely determine
the structure and properties of singularity.
It turns out that in the
presence of a vector field the oscillatory regime
exists in all spatial dimensions and for all
homogeneous models.
By analyzing the Hamiltonian equations we derive the Poincar\'e return map
associated to the Kasner indexes and fix the rules according to which the 
Kasner vectors rotate.
In correspondence to a 4-dimensional space time, the oscillatory regime here 
constructed overlap the usual Belinski-Khalatnikov-Liftshitz one.

\end{abstract}


\maketitle

\section{Introduction}

The wide interest attracted by the homogeneous cosmological models of the 
Bianchi classification \cite{BIANCHI} relies over all in the allowance for their anisotropic
dynamics. Among them the types VIII and IX stand because of their chaotic evolution
toward the initial singularity \cite{BKL70}.
The stochastic properties outlined  by such models were longly studied in order to describe 
their detailed features \cite{B82,H94,CB83,KM97P} and in recent years a new line of research has been developed 
to establish if the chaos is covariant and survives in each system of space-time coordinates \cite{M01,M03,CL97,IM01,IM02}(see also references therein. For approaches via Ashtekar variables which, providing a non-negative Hamiltonian function, could have interesting applications in the analysis of chaotic anisotropic cosmologies, see the formulation presented in \cite{Mielke92}).\newline
The cosmological interest in the Bianchi types VIII and IX universes dynamics comes out because they correspond to the maximum degree of generality allowed by the homogeneity constraint; as
 a consequence it was shown \cite{BKL82,K93,M95,BM04,BERGER} that the generic cosmological solution can be described properly, near the Big-Bang, in terms of the homogeneous chaotic dynamics as referred to each cosmological horizon (recent interest  on a special homogeneous model like the Taub one arose because in \cite{taub} was shown that it admits an accelerating stage in the dynamics as the actual Universe seems to do).\newline
It is just the dynamical decoupling of the space points which 
takes place asymptotically that allows such an inhomogeneous extension; indeed the increase of the spatial gradients is controlled by the oscillatory behavior 
of their time dependence, so that their effects are negligible at horizon size \cite{K93,M95}.
However the correspondence existing between the homogeneous dynamics and the generic inhomogeneous one holds only in 
four space-time dimensions.
In fact a generic cosmological inhomogeneous model remains characterized by chaos near the Big-Bang
  up to a ten dimensional space-time \cite{M04,DHS85,D86,HJS87,EH87} while the homogeneous models show a regular (chaos free) dynamics beyond four dimensions \cite{H02,H03}.\newline
Here we address an Hamiltonian point of view and show how the homogeneous models (of each type) single out, near the singularity, an oscillatory regime in correspondence to any number of dimensions, as soon as an electromagnetic field is included in the dynamics (for a study of the dynamics induced by an electromagnetic field in the framework of the quasi-isotropic solution, see \cite{M2000}).
In fact the presence of an electromagnetic field restores a closed potential domain for the point-universe evolution; thus, even if the considered model does not posses a sufficient numbers of potential walls to determine the oscillatory regime, its asymptotic evolution is dominated by the matter field and randomizes within the corresponding billiard
(for a connected analysis which studies the disappearance of chaos when a scalar field is included in the dynamics, though in presence of a p-form, see \cite{EWST04}).\newline
We stress how the electromagnetic billiard and its associated Poincar\'e return map coincide with  usual ones in four dimensions. However for higher dimensional spaces they  differ from the corresponding vacuum dynamics; in particular 
the map we derive outlines an interesting dependence on the number of dimensions.
A relevant out coming of our analysis consists of fixing the law for the rotation of the n-independent oscillating directions.\newline
The procedure here followed calls attention to be extended to more general contexts like those studies on the appearance of chaos in superstring dynamics \cite{DH1,DH2}.
More precisely, in Section 2 we review the Hamiltonian formalism for the General Relativity; in Section 3 we derive the Kasner solution for homogeneous models. In Section 4 we introduce the Kasner parametrization in our dynamical scheme.
Finally in Section 5 we derive the solution in the Kasner approximation, and in Section 6 we show how this dynamical system can be described in terms of the return map for the Kasner indexes and the law for Kasner vectors rotation.
In Section 7 brief Concluding remarks follow.

\section{Hamiltonian formulation}

We start by reviewing the dynamical framework of our analysis. 

Let us consider a vector field
$A_{\mu }=\left( \varphi ,A_{\alpha }\right) $
,($\alpha =1,2,\ldots ,n$)
and adopt for the metric the standard ADM 
representation \cite{ADM64,MTW}

\begin{equation}
ds^{2}=N^{2}dt^{2}-g_{\alpha \beta }\left( dx^{\alpha }+N^{\alpha }dt\right)
\left( dx^{\beta }+N^{\beta }dt\right) .  \label{met}
\end{equation}

Then the action, which describe the dynamics of the model,
takes the form

\begin{eqnarray}
 I=\int d^{n}xdt\left\{ \Pi ^{\alpha \beta }\frac{\partial }{\partial t} 
g_{\alpha \beta }+\pi ^{\alpha }\frac{\partial }{\partial t}A_{\alpha }
+\varphi D _{\alpha }\pi ^{\alpha }-NH_{0}-N^{\alpha }H_{\alpha
}\right\} ,  \label{act}
\end{eqnarray}

where

\begin{eqnarray}
 H_{0}=\frac{1}{\sqrt{g}}\left\{ \Pi _{\beta }^{\alpha }\Pi _{\alpha }^{\beta
}-\frac{1}{n-1}\left( \Pi _{\alpha }^{\alpha }\right) ^{2}+\frac{1}{2} 
g_{\alpha \beta }\pi ^{\alpha }\pi ^{\beta }+g\left( \frac{1}{4}F_{\alpha
\beta }F^{\alpha \beta }-R\right) \right\} ,  \label{hamcn}
\end{eqnarray}

\begin{equation}
H_{\alpha }=-\nabla _{\beta }\Pi _{\alpha }^{\beta }+\pi ^{\beta }F_{\alpha
\beta },  \label{momcn}
\end{equation}

denote respectively the super-Hamiltonian and super-momentum; 
here
$F_{\alpha \beta }\equiv\partial _{\beta}A_{\alpha} -\partial _{\alpha }A_{\beta }$ is the electromagnetic tensor,  $g\equiv det(g_{\alpha\beta})$ is  the determinant of the n-metric, $R$ is the n-scalar of curvature constructed by  
the metric $g_{\alpha \beta }$ and $D_\alpha\equiv \partial_\alpha+A_\alpha$.\newline
$\pi^\alpha$ and $\Pi^{\alpha\beta}$ are the conjugate momenta of the electromagnetic field and of the n-metric respectively ; they result to be a vector and a tensorial density respectively of weight $1/2$ because their expressions contain $\sqrt g$, and are defined via the relations:
\begin{equation}
\label{momento della 3 geometria}
\Pi^{\alpha \beta}=\frac{\sqrt{-g}}{N} (K^{\alpha\beta}-g^{\alpha\beta}Tr(K))
\end{equation}
\begin {equation}
\label{momento del campo elettromagnetico}
\pi^\alpha=\frac{\sqrt{-g}}{N}\left(\frac {\partial A_\beta}{\partial t} g^{\alpha \beta}-N^\beta F_\beta ^{\phantom{\beta}\alpha}\right)
\end{equation}

(here $K^{\alpha\beta}$ denotes the extrinsic curvature in the synchronous frame)

When varying this action with respect to the lapse function $N$ we obtain the Super-Hamiltonian constraint 
\begin{equation}
\label{ho}
H_0=0\,,
\end{equation}

 while its variation with respect to $\varphi$ provides the constraint $\partial _{\alpha }\pi ^{\alpha }=0$.
Since we are dealing with an abelian vector field (i.e. corresponding to an abelian group of symmetry like an electromagnetic field does) whose sources
(charged particles) are absent,
it is enough to consider only the transverse (or Lorentz) 
components for $A_{\alpha }$ and $\pi ^{\alpha }$. 
Therefore, we take the gauge conditions 
$\varphi =0$ and $D _{\alpha }\pi ^{\alpha }=0$ and this will be enough to exclude
the longitudinal parts of the vector field from the action.\newline
It is worth noting how, 
in the general case, i.e.
either in presence of the sources, either in the case of
non-abelian vector fields,
this simplification can no longer take place 
in such explicit form and, therefore, we have to
retain the term $\varphi (\partial _{\alpha }+A_\alpha)\pi ^{\alpha }$
in the action principle.\newline 
In what follows we consider the behavior
of homogeneous cosmological models in the
asymptotic limit toward the initial singularity.

\section{Homogeneous cosmological models: basic equations}

When considering the homogeneity
constraint, the whole spatial dependence of the models
can be integrated out from the action and
the dynamical variables become
only time dependent.
In general neglecting the  total Hamiltonian the spatial derivatives contained in 
$F_{\alpha \beta }$ and $R$, 
is equivalent to the so-called (generalized)
Kasner approximation.

When going over the homogeneous case, 
we choose the gauge $N^{\alpha }=0$
and, within the Kasner approximation, we get 
equations of motion for the vector field having the form 

\begin{equation}
E_{\alpha }=\frac{\partial }{\partial t}A_{\alpha }=\frac{N}{\sqrt{g}} 
g_{\alpha \beta }\pi ^{\alpha },
\end{equation}

\begin{equation}
\frac{\partial }{\partial t}\pi ^{\alpha }=0.
\end{equation}

The field equations which describe
the n-metric dynamics read as follows

\begin{equation}
\frac{\partial }{\partial t}g_{\alpha \beta }=\frac{2N}{\sqrt{g}}\left\{ \Pi
_{\alpha \beta }-\frac{1}{n-1}g_{\alpha \beta }\Pi _{\gamma }^{\gamma
}\right\} ,  \label{eq1}
\end{equation}

\begin{equation}
\frac{\partial }{\partial t}\Pi _{\beta }^{\alpha }=-\frac{N}{2\sqrt{g}}\pi
^{\alpha }\pi _{\beta }.  \label{eq2}
\end{equation}

This dynamical scheme is completed by adding to the above Hamiltonian equations the Super-Hamiltonian constraint (\ref{ho}).

\section{Kasner parameterization}

To develop the below analysis, it turns out very
convenient to adopt the so-called 
Kasner parameterization which is based on the metric 
and conjugate momentum decomposition along spatial n-bein:

\begin{equation}
g_{\alpha \beta }=\delta _{ab}l_{\alpha }^{a}l_{\beta }^{b},\,\;\ \ \ \
\;\Pi _{\alpha \beta }=p_{ab}l_{\alpha }^{a}l_{\beta }^{b},  \label{kp}
\end{equation}

where the n-bein is chosen in such a way that
the matrix $p_{ab}=diag\left(
p_{1},\ldots ,p_{n}\right) $.(for a discussion of this hamiltonian structure in correspondence to an inhomogeneous multidimensional model see \cite{KM95}); this diagonal form of the conjugate momenta is a consequence of requiring the canonical nature of the adopted transformation.
We also define a dual basis
$L_{a}^{\alpha }=g^{\alpha \beta }l_{\beta }^{a}$, such that 
$L_{a}^{\alpha }l_{\alpha }^{b}=\delta_{a}^{b}$ and $L_{a}^{\alpha }l_{\beta }^{a}=\delta_{\beta}^{\alpha}$.\newline
To face our goal we have to project (\ref{eq1}) along the Kasner vectors defined in (\ref{kp}); in this way we get the following dynamical system:

\begin{equation}
\left( L_{a}\frac{\partial }{\partial t}l_{a}\right) =\frac{N}{\sqrt{g}} 
\left( p_{a}-\frac{1}{n-1}\sum_{b}p_{b}\right) ,
\end{equation}

\begin{equation}
\left( L_{a}\frac{\partial }{\partial t}l_{b}\right) +\left( L_{b}\frac{\partial }{\partial t}l_{a}\right) =0,\;\;\;\;\;\; a\neq b.  \label{rt}
\end{equation}

Here
$\left( L_{a}l_{b}\right) =L_{a}^{\alpha }l_{b\alpha }$
denotes the ordinary vector product, treating the vector components as Euclidean ones.\\
In close analogy with above, from equation
(\ref{eq2}), we find the additional equations

\begin{equation}
\frac{\partial }{\partial t}p_{a}=-\frac{N}{2\sqrt{g}}\lambda
_{a}^{2},\;\;\;\;\;\;\;\;\lambda _{a}=\left( \pi ^{\alpha }l_{\alpha }^{a}\right) 
\label{PEQ}
\end{equation}

\begin{equation}
\left( L_{b}\frac{\partial }{\partial t}l_{a}\right) =-\frac{N}{2\sqrt{g}} 
\frac{\lambda _{a}\lambda _{b}}{p_{a}-p_{b}},\;\;\;\;\;\;\; a\neq b.  \label{eqv}
\end{equation}
In particular, we see that (\ref{eqv}) already contains (\ref{rt}).\newline
By combining together both such systems,
we obtain (\ref{PEQ}) as the first independent
equation; the equation for the Kasner vectors
takes place in the form (for the sake of simplicity we neglect the vector index
$\alpha $)

\begin{equation}
\frac{\partial }{\partial t}l_{a}=\frac{N}{\sqrt{g}}\left\{ \left( p_{a}- 
\frac{1}{n-1}\sum_{b}p_{b}\right) l_{a}-\frac{1}{2}\sum_{\;b\neq a.}\frac{ 
\lambda _{a}\lambda _{b}}{p_{a}-p_{b}}l_{b}\right\} .  \label{VEQ}
\end{equation}

We want to put in evidence the oscillatory regime that the bein vectors possess; to better analyze this dynamics let us distinguish scale functions in the following way

\begin{equation}
\label{riscalate}
l_{a}=\exp \left( q^{a}/2\right) \ell _{a}.
\end{equation}
\begin{equation}
L_{a}=\exp \left( -q^{a}/2\right) {\cal L} _{a}.
\end{equation}

being $\ell_a$ the so called Kasner vectors.

Thus instead of (\ref{PEQ}) and (\ref{VEQ}) we rewrite 

\begin{equation}
\frac{\partial }{\partial t}p_{a}=-\frac{N}{2\sqrt{g}}\widetilde{\lambda } 
_{a}^{2}\exp \left( q^{a}\right) ,\;\;\;\;\;\;\;\widetilde{\lambda }_{a}=\left( \pi
^{\alpha }\ell _{\alpha }^{a}\right) ,  \label{mixm2}
\end{equation}

\begin{equation}
\frac{\partial }{\partial t}q^{a}=\frac{2N}{\sqrt{g}}\left( p_{a}-\frac{1}{ 
n-1}\sum_{b}p_{b}\right) ,  \label{mixm1}
\end{equation}

\begin{equation}
\frac{\partial }{\partial t}\ell _{a}=-\frac{N}{2\sqrt{g}}\sum_{\;b\neq a.} 
\frac{\widetilde{\lambda }_{a}\widetilde{\lambda }_{b}}{p_{a}-p_{b}}\exp
\left( q^{b}\right) \ell _{b}=\sum_{\;b\neq a.}\frac{\widetilde{\lambda } 
_{a}\left( \frac{\partial }{\partial t}p_{b}\right) }{\widetilde{\lambda } 
_{b}\left( p_{a}-p_{b}\right) }\ell _{b}.  \label{rot}
\end{equation}

The last two equations are obtained by substituting variables (\ref{riscalate}) and then projecting (\ref{VEQ}) on the vectors ${ L}^\beta_c$; when the index $c$ is equal to $a$, we get (\ref{mixm1}), otherwise we get (\ref{rot}) (in which we neglected a term vanishing in the exact Kasner-like solution, and therefore, in our analysis, of higher order).  

To this system we should also add the Hamiltonian constraint

\begin{equation}
H_{0}=0=\frac{N}{\sqrt{g}}\left\{ \sum p_{a}^{2}-\frac{1}{n-1}\left( \sum
p_{a}\right) ^{2}+\frac{1}{2}\sum e^{q_{a}}\widetilde{\lambda } 
_{a}^{2}\right\} .  \label{Hconstr}
\end{equation}

To require that the quantities $\widetilde{\lambda }_{a}$ are constants implies that vectors $\ell^a$ do not depend on time in turn and therefore no Kasner vectors rotation takes place (see section 6); such a situation corresponds exactly to the n-dimensional Kasner behavior \cite{DHS85,D86,LL}. Since the rotation of Kasner vectors is induced by an higher order term for $g\rightarrow 0$ with respect to the pure Kasner dynamics, we can assume that $\widetilde{\lambda }_{a}$ are near to be constant and therefore near the singularity (\ref{mixm2}), (\ref{mixm1}) give, in the asymptotic limit to the cosmological singularity, the billiards \cite{IM01,BM04,B82} on $\left( n-1\right) -$dimensional Lobachevsky space, exactly like in the 3-dimensional  mixmaster case \cite{M69}. By other words, in our scheme the evolution is not simply Kasner like, but we get a dynamical picture in which the {\em point Universe} moves according to a piecewise Kasner solution.
The features predicted by the two equations (\ref{mixm2}) and (\ref{mixm1}) will be discussed in section 5 and 6.

We remark that the validity of the Kasner approximation is based on the possibility to neglect the n-dimensional Ricci scalar with respect to the term containing time derivatives. In the Kasner solution this picture holds along the whole system evolution to the initial singularity \cite{LL}; instead in the piecewise Kasner solution the n-dimensional Ricci tensor is negligible only for finite time intervals ending with a bounce against the potential terms arising from the spatial curvature \cite{BKL70}.
However in both these cases the validity of the Kasner approximation is confirmed by a large number of theoretical and numerical investigations, concerning homogeneous and inhomogeneous cosmological models near the singularity \cite{ BKL70,B82,H94,CB83,KM97P,IM01,BKL82,K93,BM04,BERGER,M69}

Our scheme relies on the choice of a synchronous (or Gaussian) reference frame which corresponds to have $N=1$ and $N^\alpha=0$. This system has many interesting properties, among which out stands its geodesic nature \cite{LL}; in fact the line $x^\alpha=constant$ results to be geodesic of the manifold and since the normal to spatial hypersurfaces reads $n^\mu=(1,\vec 0)$ free particles are co-moving to this system.
>From a cosmological point of view the Gaussian frame is relevant because the galaxies are (almost) free particles in the actual Universe and therefore our phenomenology relies on a synchronous gauge. However the results obtained in this work have to remain valid for other gauge choices in view of the general covariance.

\section{Kasner solution}

To analyze the time evolution of the Kasner vectors and to obtain their rotation law, it result to be convenient to project them on the time independent quantities $\pi ^{\alpha }$ and to decompose them in the two parts: 
\begin{equation}
\label{123}
\vec{\ell}_{a}=\vec{\ell}_{a\parallel }+\vec{\ell}_{a\perp };\;\;\;\;\;\;\vec{\ell} 
_{a\parallel }=\frac{\widetilde{\lambda }_{a}}{\pi ^{2}}\vec{\pi},\;\;\;\;\;\;\left( 
\vec{\pi}\vec{\ell}_{a\perp }\right) =0.
\end{equation}
Here, for the sake of simplicity, we use the vector notation which is useful when treating the components of these vectors as Euclidean ones. 

Hence we can split (\ref{rot}) into the two independent components 
\begin{equation}
\frac{\partial }{\partial t}\widetilde{\lambda }_{a}=\sum_{\;b\neq a.}\frac{ 
\left( \frac{\partial }{\partial t}p_{b}\right) \widetilde{\lambda }_{a}}{ 
\widetilde{\lambda }_{b}\left( p_{a}-p_{b}\right) }\widetilde{\lambda } 
_{b}\;,  \label{Leq}
\end{equation}
\begin{equation}
\frac{\partial }{\partial t}\vec{\ell}_{a\perp }=\sum_{\;b\neq a.}\frac{ 
\left( \frac{\partial }{\partial t}p_{b}\right) \widetilde{\lambda }_{a}}{ 
\widetilde{\lambda }_{b}\left( p_{a}-p_{b}\right) }\vec{\ell}_{b\perp
}=A_{b}^{a}\left( t\right) \vec{\ell}_{b\perp }\;.  \label{A}
\end{equation}
Since above the matrix $A_{b}^{a}$ does not depend on $\vec{\ell}_{a\perp }$, 
then we get the formal solution 
\begin{equation}
\vec{\ell}_{a\perp }\left( t\right) =T\exp \left\{
\int\limits_{t_{0}}^{t}A_{b}^{a}\left( t^{\prime }\right) dt^{\prime
}\right\} \vec{\ell}_{b\perp }\left( t_{0}\right) ,
\end{equation}
The remaining equations (\ref{Leq}) together with (\ref{mixm2}), (\ref{mixm1}), and (\ref
{Hconstr}) provide a self-consistent dynamical system.

Since the Kasner solution corresponds to neglect the contributions of the n-dimensional Ricci scalar to the dynamics, then to get such a behavior we have to take the limit in which all the terms $\exp \left(
q^{a}\right)$ become of higher order;
under this assumption we get the following simplified dynamical system

\begin{equation}
\begin{array}{c}
p_{a}=const,\;\;\;\;\;\;\;\;\;\;\;\widetilde{\lambda }_{a}=const,\;\;\;\;\;\;\;\;\;\;\;\;\vec{\ell}_{a\perp
}=const, \\ 
\frac{\partial }{\partial t}q_{a}=\frac{2N}{\sqrt{g}}\left( p_{a}-\frac{1}{ 
n-1}\sum_{b}p_{b}\right) , \\ 
\sum p_{a}^{2}-\frac{1}{n-1}\left( \sum p_{a}\right) ^{2}+\frac{1}{2}\sum
e^{q_{a}}\widetilde{\lambda }_{a}^{2}=0\;,
\end{array}
\end{equation}
whose solution, in the gauge $N=1$ and toward the cosmological singularity ($g\rightarrow 0$), takes the form 
\begin{equation}
g_{\alpha \beta }=\sum_{a}t^{2s_{a}}\ell _{\alpha }^{a}\ell _{\beta
}^{a}\;,\;\;\;\;\;\;\;\;\;\;\;\;\;\;\;s_{a}=1-\left( n-1\right) \frac{p_{a}}{\sum_{b}p_{b}}\;,
\end{equation}
where the Kasner indexes  $s_{a}$ satisfy the identities 
\begin{equation}
\sum s_{a}=\sum s_{a}^{2}=1.
\end{equation}
Let's take the Kasner indexes in the increasing order 
\begin{equation}
s_{1}\leq s_{2}\leq \ldots \leq s_{n};
\end{equation}
in this way we always have $s_{1}<0$ and $s_{n}\geq s_{n-1}\geq 0$. Therefore the
largest increasing term (as $t\rightarrow 0$
 $t^{s_{1}}\rightarrow \infty $) among the neglected ones comes from $s_{1}$  and it is to be taken into account to construct the oscillatory regime toward the cosmological singularity.

\section{Billiard representation: the return map and the rotation of Kasner vectors}

To construct the oscillatory regime we retain just the leading term corresponding to $\exp \left( q^{1}\right) $, so the field equations (\ref{mixm2}) and (\ref{mixm1}) rewrite 
\begin{eqnarray}
\label {28}
\frac{\partial }{\partial t}\widetilde{\lambda }_{1} &=&0\;,  \nonumber \\
\frac{\partial }{\partial t}\widetilde{\lambda }_{a} &=&\frac{\left( \frac{ 
\partial }{\partial t}p_{1}\right) \widetilde{\lambda }_{a}}{\left(
p_{a}-p_{1}\right) }\;,\;\;\;\; a\neq 1,  \nonumber \\
\frac{\partial }{\partial t}p_{a} &=&0,\;a\neq 1, \\
\frac{\partial }{\partial t}p_{1} &=&-\frac{N}{2\sqrt{g}}\widetilde{\lambda } 
_{1}^{2}\exp \left( q^{1}\right) ,  \nonumber \\
\frac{\partial }{\partial t}q_{a} &=&\frac{2N}{\sqrt{g}}\left( p_{a}-\frac{1 
}{n-1}\sum_{b}p_{b}\right) .  \nonumber
\end{eqnarray}
The first of equations (\ref{28}) gives $\widetilde{\lambda }_{1}=const$, while the second admits
the solution 
\begin{equation}
\widetilde{\lambda }_{a}\left( p_{a}-p_{1}\right) =const.
\end{equation}
The remaining part of the dynamical system allows us to determine the return map governing the replacements of Kasner epochs (i.e. intervals of time during which the evolution is Kasner like) 
\begin{equation}
s_{1}^{\prime }=\frac{-s_{1}}{1+\frac{2}{n-2}s_{1}},\;\;\;\;\;\;\;\;\;\;s_{a}^{\prime }= 
\frac{s_{a}+\frac{2}{n-2}s_{1}}{1+\frac{2}{n-2}s_{1}},
\end{equation}
\begin{equation}
\label{234}
\widetilde{\lambda }_{1}^{\prime }=\widetilde{\lambda }_{1},\;\;\;\;\;\;\;\;\;\;\widetilde{ 
\lambda }_{a}^{\prime }=\widetilde{\lambda }_{a}\left( 1-2\frac{\left(
n-1\right) s_{1}}{\left( n-2\right) s_{a}+ns_{1}}\right) ,
\end{equation}

Defining the quantities $k_{a}=p_{a}-\frac{1}{n-1} 
\sum p_b$ we find for them the following iteration law
\begin{eqnarray}
k_{1}^{\prime } &=&-k_{1} \\
k_{a}^{\prime } &=&k_{a}+\frac{2}{n-2}k_{1} \\
\sum k^{\prime } &=&\sum k+\frac{2}{n-2}k_{1}
\end{eqnarray}

Our analysis is completed by investigating the rotation of Kasner vectors ${\vec{\ell}_a}$ through the epochs replacements; by (\ref{A}) we get the following system 
\begin{equation}
\label{345}
\frac{\partial }{\partial t}\vec{\ell}_{a\perp }=\frac{\left( \frac{\partial 
}{\partial t}p_{1}\right) \widetilde{\lambda }_{a}}{\widetilde{\lambda } 
_{1}\left( p_{a}-p_{1}\right) }\vec{\ell}_{1\perp },\,\;\;\;\;\;\;\;\frac{\partial }{ 
\partial t}\vec{\ell}_{1\perp }=0,
\end{equation}
admitting the integral 
\begin{equation}
\widetilde{\lambda }_{1}\vec{\ell}_{a\perp }-\widetilde{\lambda }_{a}\vec{ 
\ell}_{1\perp }=const.
\end{equation}
Putting together (\ref{123}), (\ref{234}) and (\ref{345}) we arrive to the final iteration law
\begin{eqnarray}
\label{39}
\vec{\ell}_{a}^{\prime }=\vec{\ell}_{a}+\sigma _{a}\vec{\ell}_{1},\;\;\;\;\;\;\;\;\sigma
_{a}& =\frac{\widetilde{\lambda }_{a}^{\prime }-\widetilde{\lambda }_{a}}{ 
\widetilde{\lambda }_{1}}=\nonumber\\
&=-2\frac{\left( n-1\right) s_{1}}{\left( n-2\right)
s_{a}+ns_{1}}\frac{\widetilde{\lambda }_{a}}{\widetilde{\lambda }_{1}}.
\end{eqnarray}
which completes our dynamical scheme.

Thus the homogeneous Universes here discussed approaches the initial singularity being described by a metric tensor with oscillating scale factors and rotating Kasner vectors.
Passing from one Kasner epoch to another one, the negative Kasner index $s_1$ is exchanged between different directions (for istance $\vec\ell_1$ and $\vec\ell_2$) and at the same time this directions rotate in the space according to the law (\ref{39}).
The presence of a vector field is crucial because, independently on the considered model, it induces a (dynamically \cite{BM04}) closed domain on the configuration space.
The amplitude of $q^a$-oscillations increase approaching the initial singularity and  their minimum value approaches $-\infty$

In correspondence of this oscillation of the scale factor the Kasner vectors $\vec{\ell}_{a}$ rotate and, at the lowest order in $q^a$, the quantities $\sigma
_{a}$ remain constant along a Kasner epoch; in this sense the vanishing behavior of the determinant $g$ approaching the singularity does not affect significantly the rotation law (\ref{39}). We conclude by stressing that duration of a Kasner epoch decrease as the singularity is approached in this scheme.

\section{Concluding remarks}

By the study above developed, we have shown how any multidimensional homogeneous cosmological model
 acquires, near the singularity, an oscillatory regime when an electromagnetic field is included in the dynamics.
This result is a consequence of the capability that a vector field has to generate a billiard configuration in the
 asymptotic evolution; such a "billiard-ball" representation of the universe dynamics coincides with the BKL (Belinski-Khalatnikov-Liftshitz) piecewise approach only in the 4-dimensional space-time while in higher dimensions new features appear. In particular the obtained oscillatory regime characterizes all the homogeneous models, disregarding their potential term; furthermore the map by which the Kasner indexes evolve acquires a direct dependence on the dimensions number.

A valuable issue of our Hamiltonian dynamics relies on fixing the rule of the Kasner vectors rotation.
Such a law of rotation is relevant to connect the Cauchy problem with later stages of the system dynamics.
More precisely the rotation of the Kasner vectors is sensitive to the boundary conditions on the "matter" fields and has to be taken into account when using the studied homogeneous models within a cosmological "picture".

Once the spatial gradients are taken into account, our result can be extended to a generic inhomogeneous cosmological model, in the same spirit as the Bianchi type VIII and IX model oscillatory regime is upgraded in four dimensions.

Such an extension will reliably show how, in the presence of a vector field, the generic cosmological solution is described, in correspondence to any dimensions number, by an oscillatory approach to the Big-Bang.

Investigations in this direction, as well as extended to the more general frameworks of superstrings \cite{DH1,DH2} and brane dynamics \cite{Ma04} have to be subject for further developments.

\end{document}